\documentclass[12pt]{article}

\usepackage{graphicx}
\usepackage{epsfig}
\usepackage{amsfonts}
\usepackage{amssymb}

\newcommand{\ee}{\end{equation}}
\newcommand{\eea}{\end{eqnarray}}
\newcommand{\be}{\begin{equation}}
\newcommand{\bea}{\begin{eqnarray}}

\newcommand{\re}[1]{(\ref{#1})}

\begin{document}

\title{
Stable  black hole
solutions with non-Abelian fields}
 \vspace{1.5truecm}
\author{ 
{\large Eugen Radu}$^{\dagger}$ 
and {\large D. H. Tchrakian}$^{\star \diamond }$ 
\vspace*{0.2cm}
\\
$^{\dagger}${\small Institut f\"ur Physik, Universit\"at Oldenburg,
D-26111 Oldenburg, Germany} 
   \\
$^{\star}${\small  Department of Computer Science,
National University of Ireland Maynooth,
Maynooth,
Ireland} \\
$^{\diamond}${\small School of Theoretical Physics -- DIAS, 10 Burlington
Road, Dublin 4, Ireland }} 
\date{\today}
\maketitle

\begin{abstract}
We construct finite mass, asymptotically flat black hole solutions in $d=4$ 
Einstein--Yang-Mills theory augmented with higher order curvature terms of the gauge field.
They possess non-Abelian hair in addition to Coulomb electric charge, and,
below some non-zero critical temperature, they are thermodynamically preferred over the 
Reissner-Nordstr\"om solution. 
Our results indicate the existence of hairy non-Abelian black holes which are stable 
under linear, spherically symmetric perturbations.
\end{abstract}

\noindent{\textbf{~~~Introduction.--~}}
In recent years it has been realized that the electrically charged Reissner-Nordstr\"om (RN) black hole,
when considered as solution of a more general theory, may become unstable to forming hair at low temperatures. 
This has lead to the discovery of some holographic models for condensed matter systems, and, in particular, to a
gravitational description of superconductivity (see \cite{Horowitz:2010gk} for a review).
 
The case of Einstein-Yang-Mills (EYM) model with negative cosmological constant $\Lambda$ in $d=4$ spacetime dimensions provides an
interesting illustration of these aspects.
As shown in \cite{Gubser:2008zu}, there is a second order phase transition
between the RN--anti-de Sitter solutions, which are preferred at high temperatures, and
symmetry breaking non-Abelian black holes, which are preferred at low temperatures. 
 In \cite{Gubser:2008zu}, $\Lambda$ plays an essential role;
although  electrically charged hairy black holes
do exist
 also in a Minkowski spacetime background \cite{Galtsov:1991au}, they have rather different properties
as compared to the anti-de Sitter (AdS) solutions in \cite{Gubser:2008zu}.
In particular they do not emerge as perturbation
of the RN black holes, and, similar to the well-known $d=4$ asymptotically flat, purely magnetic
EYM solutions \cite{Volkov:1998cc}, are also perturbatively unstable.

However, one might take the view that in the strong coupling regime 
the ($\Lambda=0$) EYM theory
is incomplete.
Perhaps the simplest possibility to describe this situation is to 
supplement the action of the EYM model with higher order curvature terms, 
for both gravitational and gauge field sectors.
As discussed $e.g.$ in \cite{Donets:1995ya}, the inclusion of (string theory inspired-) corrections to the gravity 
action does not lead to qualitatively new features. By contrast, we will here
argue that the situation is different {\it vis a vis} the inclusion of higher order Yang-Mills (YM) curvature
terms.
 This possibility
has been overlooked so far in the literature.
The first relevant order in this case is the fourth, in which case the most general such density added to
the Lagrangian  
consists 
of the four terms,
\begin{eqnarray}
\label{Ls}
&&\mathcal{L}_s=
c_1 {\rm Tr}\left\{  F_{\mu\nu}F_{\rho\sigma} F^{\mu\nu}F^{\rho\sigma} \right \}
+
c_2 {\rm Tr}\left\{  F_{\mu\nu}F^{\mu\nu}  F_{\rho\sigma}F^{\rho\sigma} \right \}
\\
&&{~~~~~~~~~}+
c_3 {\rm Tr}\left\{  F_{\tau\nu}F^{\mu\tau}  F_{\mu\lambda}F^{\lambda\nu} \right \}
+c_4 {\rm Tr}\left\{  F_{\mu\nu}F^{\nu\rho}  F_{\rho\lambda}F^{\lambda\mu} \right \},
\nonumber
\end{eqnarray}
with some constant coefficients $c_i$.
 A particularly priviledged such combination,   which we adopted here, is that with
$c_1=c_2=-4 c_3,$ $c_4=0$.
In that case,  $\mathcal{L}_s$ features only the second power of any ``velocity field'' and
is a causal density just
like the Gauss-Bonnet term in gravity \cite{Zwiebach:1985uq} or the
Skyrme~\cite{Skyrme:1962vh}
term of the $O(4)$ sigma model.
With this specific choice of the constants $c_i$,
the Lagrange density
\re{Ls} is nothing else than the trace of the square of the $4-$form curvature
  $F_{\mu\nu \rho \sigma}= \{F_{\mu [\nu},F_{\rho\sigma]} \}$.
This is  the  
second member of the YM hierarchy \cite{Tchrakian:1984gq},
providing a natural generalization of the usual YM model. 
 A convenient way to express this system  
is  $ {\rm Tr}\left\{ (F_{\mu \nu} {}\tilde F^{\mu \nu})^2 \right \}$, 
where a tilde denotes the Hodge dual.
 
Notwithstanding our specific choice for the constants $c_i$ in (\ref{Ls}), 
we have verified that
for certain other choices, some salient
features of the solutions
discussed in this work, in particular
the instability of the RN black hole, persist. 

\noindent{\textbf{~~~The model.--~}}
Ignoring for simplicity other possible corrections, 
we consider the following action for the model
\begin{eqnarray}
\label{action}
S=\int d^4 x 
\sqrt{-g}
\bigg [
\frac{1}{4}R-\frac{1}{2 }{\rm Tr}\left \{ F_{\mu\nu}F^{\mu\nu} \right\}
+\frac{3\tau}{2 }
 {\rm Tr}\left\{ (F_{\mu \nu} {}\tilde F^{\mu \nu})^2 \right \}\bigg ],
\end{eqnarray}
(here we have set $4\pi G/e^2=1$, such that the only parameter of the theory is $\tau$).

In what follows, we shall prove that the presence of the last term in (\ref{action})
leads to an instability of the RN black hole, together with the occurance of stable black holes
with non-Abelian hair outside the horizon.
We shall restrict
attention to the following spherically symmetric Ansatz: 
 \begin{eqnarray}
\label{metric}
ds^2=\frac{dr^2}{N}+r^2(d\theta^2+\sin^2 \theta d\phi^2)-N\sigma^2dt^2,~
\end{eqnarray} 
where $N,\sigma$ are functions of  $ r$ and $t$ in general.
The minimal gauge group for which the
superposition of a Coulomb field and a non-Abelian hair is not forbidden by the 'baldness'
theorems \cite{bald}  is $SU(3)$.
Then, as in the $\tau=0$ case in \cite{Galtsov:1991au}, 
we shall restrict  to an $SU(2)\times U(1)$ truncation 
of the $SU(3)$ group,  
the general spherically symmetric ansatz for the gauge potential being
\begin{eqnarray}
\label{YMansatz}
A= 
  \bigg \{
 (\nu T_3+U T_8) dr+
 (w T_1+\tilde w T_2) d\theta
 +\left( (w T_2-\tilde w T_1)\sin \theta +\cos \theta T_3 \right) d\phi
 +(v T_3 +V T_8) dt
 \bigg \} 
 ,
\end{eqnarray}
where $\nu,w,\tilde w,v$ and $U,V$ are functions of $(r,t)$
and $T_i$ are the standard generators of the $SU(3)$ Lie algebra.

For static solutions, 
one can set the functions $\nu,\tilde w, v$ and $U$ to zero without any loss of generality,
resulting in the equations
\begin{eqnarray}
\nonumber
&&m'=Nw'^2+\frac{(1-w^2)^2}{2r^2}+\frac{r^2 V'^2}{2\sigma^2}+\tau \frac{(1-w^2)^2V'^2}{r^2 \sigma^2},
~~~
\sigma'=\frac{2\sigma}{r}w'^2,
\\
\label{eqs}
&&w''+(\frac{N'}{N}+\frac{\sigma'}{\sigma})w'
+\frac{w(1-w^2)}{r^2 N}+\frac{2\tau (w^2-1)V'^2}{r^2N\sigma^2}=0,
 \end{eqnarray}
 together with the first integral for the electric potential,  
\begin{eqnarray}
\label{int-V}
 V'=Q\frac{\sigma}{r^2}\left(1+\frac{2\tau(1-w^2)^2}{r^4}\right)^{-1},
\end{eqnarray}
with $Q$ an arbitrary constant.

The RN solution corresponds to a vanishing SU(2) field, $w(r)=\pm 1$,
and  $m(r)=M-\frac{Q^2}{2r}$, $\sigma=1$, $V(r)=\Phi-Q/r$.
This  solution has an outer event horizon at $r_h=M+\sqrt{M^2-Q^2}$,
which becomes extremal for $Q= M$.
 
Solutions with 
nonzero magnetic gauge fields should also exist.
 However, one can see that, for  $Q\neq 0$,
  the
first integral (\ref{int-V})  excludes the existence of particle-like
configurations with a regular origin. 
 Thus, the only physically interesting solutions 
 of this model describe black holes,
 with an event horizon at $r=r_h>0$, located at the largest root of $N(r_h)=0$.
The regularity conditions at the horizon imply the following series expansion there  
\begin{eqnarray}
\nonumber
&&m(r)=\frac{r_h}{2}+m_1(r-r_h)+\dots,
~
\sigma(r)=\sigma_h+\frac{2 \sigma_h w_1^2}{r_h}(r-r_h)+\dots,
\\
\label{exp-eh}
&&w(r)=w_h+w_1(r-r_h)+\dots,~
 V(r)=v_1(r-r_h)+\dots,
\end{eqnarray} 
where
$
v_1=\frac{Q r_h^2 \sigma_h}{r_h^4+2\tau (1-w_h^2)^2},
$
$
m_1=\frac{ \sigma_h^2(1-w_h^2)^2+v_1^2(r_h^4+2\tau (1-w_h^2)^2)}{2r_h \sigma_h^2}
$,
$
w_1= \frac{(\sigma_h^2-2\tau v_1^2)w_h(w_h^2-1)}
{(1-2 m_1)r_h\sigma_h^2}.
$
It is also straightforward to show that the requirement of finite energy 
implies the following asymptotic behavior at large $r$ 
\begin{eqnarray}
\label{exp-inf}
 m(r)=M-\frac{Q^2}{2r} +\dots,
~~
\sigma(r)=1-\frac{ J^2}{2r^4}+\dots,
~~ w(r)=\pm 1+\frac{ J}{r}+\dots,~~V(r)=\Phi-\frac{Q}{r}+\dots~.
\end{eqnarray} 
Once the parameters $\sigma_h,~w_h$ and $J,~M,~Q$ are specified, all other coefficients in 
(\ref{exp-eh}), (\ref{exp-inf}) can be computed order by order.
$M$ and $Q$ correspond to the mass and electric charge of the solutions;
other quantities of interest are 
the Hawking temperature $T_H= \frac{1}{4 \pi} \sigma(r_h) N'(r_h)$, 
the entropy $S=\frac{A_H}{4}={\pi r_h^2}$ and the chemical potential $\Phi$.
$J$ here is an order parameter describing the deviation from the Abelian solution.
 
\noindent{\textbf{~~~The results.--~}}
The solutions of the equations (\ref{eqs}), (\ref{int-V})
interpolating between the  asymptotics (\ref{exp-eh}), (\ref{exp-inf})
were constructed numerically, by employing a shooting strategy.
 Finite mass, non-Abelian black holes exist 
for any $\tau \geq 0$.
For given $Q,r_h,\tau$,  the solutions are found  for discrete
 values of the parameter $w_h$, labeled by the number of nodes, $n$, of the magnetic
 YM potential $w(r)$.
To simplify the picture,
  in this work we have restricted
attention to solutions with a monotonic behavior 
of the magnetic gauge potential $w(r)$ (and thus $n=0,1$ only).

In characterizing the non-Abelian configurations, it is convenient to introduce the quantity
\begin{eqnarray}
\label{kappa}
\kappa=\frac{\tau}{Q^2}.
\end{eqnarray} 
For $0\leq \kappa< 1/2$,
the solutions can be thought of as nonlinear superpositions
of the RN and the purely magnetic SU(2) black holes in \cite{89}.
In particular, they are unstable since the magnetic gauge potential has always $n\geq 1$ nodes.
 Moreover, their free energy $F=M-T_H S$ 
is always greater than that of the RN configuration with the same $T_H$ and $Q$. 

 \begin{figure}[t]
\begin{center}
\epsfysize=6.5cm
\mbox{\epsffile{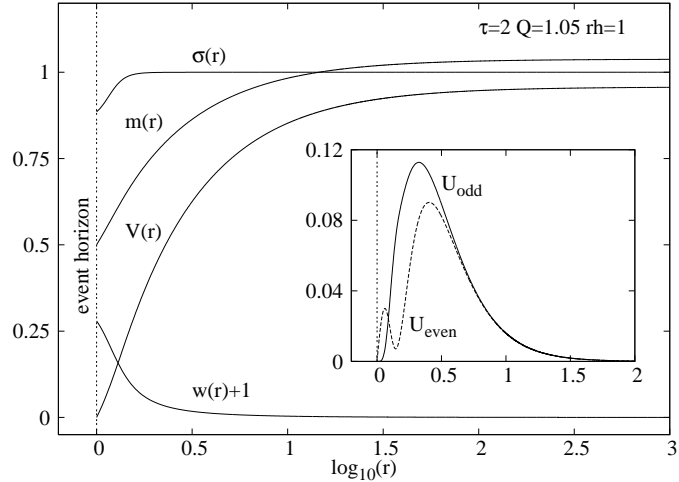}}
\caption{
The profiles of a typical non-Abelian  nodeless solution are shown
together with the corresponding potentials for the perturbation equations
(\ref{schr-even}), (\ref{schr-odd})  below.  
}
\end{center}
\end{figure}

The picture is very different for $\kappa\geq 1/2$.
In this case, for given $\kappa$, we notice the existence of a set of non-Abelian solutions
emerging as perturbations of the RN black holes,
for a critical value of the charge to mass ratio. 
This instability of the Abelian configuration is found within the Ansatz (\ref{YMansatz}),
for values of the magnetic gauge potential
$w(r)$ close to the vacuum values $ \pm 1$ everywhere, $w(r)= \pm 1+ \epsilon W(r)$.
The perturbation $W(r)$  starts from some nonzero value at the horizon and vanishes at infinity,
being a solution of the linear equation
\begin{eqnarray}
\label{stab1}
 (N W')'-(1-\frac{2\tau Q^2}{r^4})\frac{2W}{r^2}=0~,
\end{eqnarray}
where $N=(1-\frac{r_h}{r})(1-\frac{Q^2}{r_h r})$. 
The second term in this equation 
 can be seen as an effective mass term $\mu^2$ for $W$ near the horizon, with 
$\mu^2\sim 1-2\kappa (Q/r_h)^4$.  
We have found that for any $\kappa\geq 1/2$ ($i.e.$ $\mu^2<0$), an instability occurs for a critical
value of the mass to charge ratio of the RN solution. 
An approximate value of this ratio is found by 
 using an asymptotic matching expansion for the approximate solutions of (\ref{stab1})
 at the horizon and at infinity,  the result 
 \begin{eqnarray}
\label{QM} 
\frac{Q}{M}=\frac{2\sqrt{6}\sqrt{1+\sqrt{1+48\kappa}}}{7+\sqrt{1+48\kappa}} 
\end{eqnarray}
providing good agreement with the numerical data.

This unstability signals the
 emergence of a symmetry breaking branch of the non-Abelian solution
bifurcating from the RN black holes.
In contrast to the solutions with $\kappa <1/2$,
here we notice the existence of a fundamental branch of solutions without nodes
 in the magnetic gauge function $w(r)$,
 see $e.g.$ the typical profile shown in Fig. 1.

 \begin{figure}[t]
\begin{center}
\epsfysize=6.5cm
\mbox{\epsffile{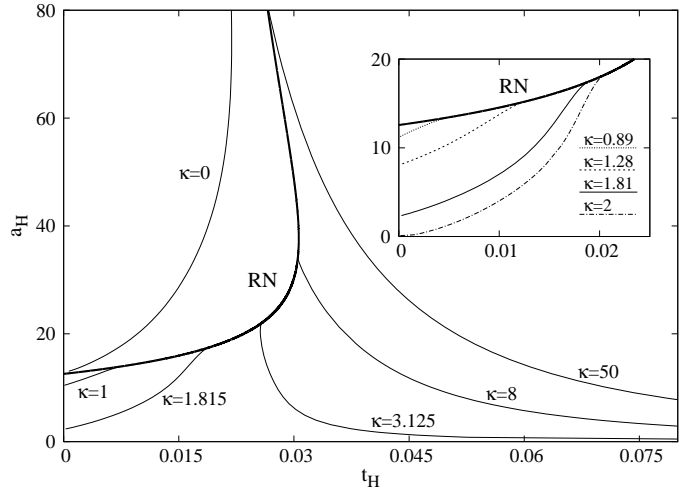}}
\caption{
The scaled horizon area $a_H=A_H/Q^2$ is plotted $vs.$ the scaled temperature 
$t_H=T_H Q$  for
several values of the  ratio $\kappa=\tau/Q^2$. 
}
\end{center} 
\end{figure}

\begin{figure}[t]
\begin{center}
\epsfysize=6.5cm
\mbox{\epsffile{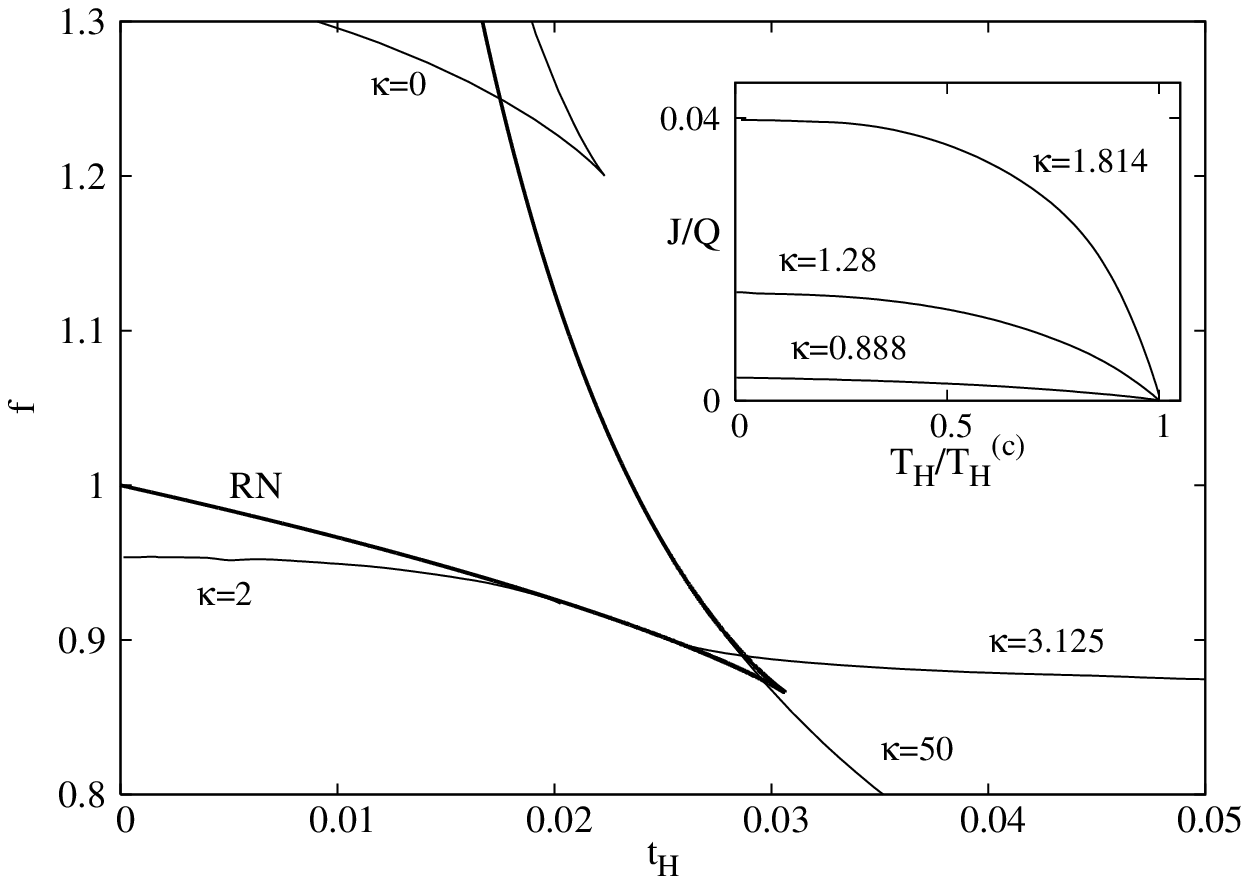}}
\caption{
The scaled free energy $f=F/Q$ is plotted $vs.$ the scaled temperature $t_H=T_HQ$ for
several values of  $\kappa=\tau/Q^2$.
The inlet shows the behavior of the order parameter $J$
which enters the asymptotics of the magnetic potential.
}
\end{center}
\end{figure}
Among the solutions bifurcating from RN black holes, those with $1/2 \leq \kappa \leq  2$
are of special interest, sharing some striking similarity with the picture found
for the EYM-AdS system in \cite{Gubser:2008zu}.
A plot of the horizon area as a function of the temperature reveals the existence 
of a single branch of non-Abelian solutions, which exist below a critical temperature
$T_H^{(c)}$ only (the range of the scaled temperature $t_H=T_H Q$ being $0<t_H<0.0203$).
In a canonical ensemble, the non-Abelian black holes within this range of
$\kappa$
exist for a finite interval of  $r_h$ ($i.e.$ of the entropy) only. 
These solutions always possess a positive specific heat, the Hawking temperature vanishing  
for a minimal value $r_h^{(min)}= \sqrt{2\tau}\sqrt{\sqrt{2\kappa}-1}$
of the event horizon radius. As $r\to r_h^{(min)}$,
an extremal non-Abelian black hole solution with a regular horizon
is approached, the charge to mass ratio of this configurations being always greater than one.
Furthermore, it turns out that the free energy 
of a RN solution is always larger than the free energy
of a non-Abelian solution with the same temperature and electric charge.
Therefore,  for $1/2 \leq \kappa \leq  2$ these non-Abelian black holes are preferred. 
The difference of free energies scales like $(T_H^{(c)}-T_H)^2$ near the
transition point,  signaling that this is a second order phase transition, while
$J \sim \sqrt{1-T_H/T_H^{(c)}}$. Moreover, for the same values of the mass and electric charge, 
the RN solution has a smaller event horizon radius (and thus a smaller entropy),
than the non-Abelian black hole.

The picture is somehow different for $\kappa>2$.
Again, one finds a single branch of solutions emerging as a perturbation of the RN black hole,
nodeless configurations existing also in this case.
However, these solutions exist for values of the temperature greater than the critical
value $T_H^{(c)}$ and are thermally unstable.

Some of these features are shown in Fig. 2 where we plot the scaled
horizon area $a_H=A_H/Q^2$
 as a function of the scaled temperature 
$t_H= T_H  Q$ for several values of the ratio $\tau/Q^2$. 
The branch of RN solutions is also shown there.
In Fig. 3 the scaled free energy $f=F/Q$  is 
plotted as a function of the scaled temperature $t_H$ for several values of $\kappa$. 
The inlet there shows the behavior of the parameter $J$ which enters the 
asymptotic behavior  of the magnetic potential, as a function of the ratio $T_H/T_H^{(c)}$.

\noindent{\textbf{~~~Existence of stable solutions.--~}}
An outstanding question now is whether the $F^4$ term leads also 
to stable non-Abelian black holes.
The fact that, for a range of $\kappa$, we have found nodeless 
solutions which are thermodynamically
favored over the RN black holes suggests a positive answer to this question. 

For simplicity, we consider linear, spherically
symmetric perturbations only. Even in this case, the analysis is
highly involved and the details will be
presented elsewhere. Here we briefly outline just the key features.

As usual, all field variables are written as the sum of the static equilibrium solution whose
stability we are investigating and a time dependent perturbation.
Choosing a gauge such that $U=v=0$, one finds that the fluctuations decouple in two groups.
$\delta w(r,t)$, $\delta V(r,t)$, $\delta \sigma(r,t)$
and $\delta m(r,t)$
form even-parity perturbations,
whereas $\delta \tilde w(r,t)$ and $\delta \nu(r,t)$
form odd-parity perturbations.
The linearized equations imply that $\delta \sigma(r,t)$,  $\delta V(r,t)$
and $\delta m(r,t)$
are determined by  $\delta w(r,t)=w_1(r)e^{-i\Omega t}$,
leading to a single Schr\"odinger equation
\begin{eqnarray}
\label{schr-even}
-\frac{d^2 w_1}{d \rho^2}+U_{even}(\rho)w_1=\Omega^2 w_1,
\end{eqnarray}
(where $dr/d \rho=N \sigma $) with a potential
\begin{eqnarray}
\label{pot-even}
&&
 U_{even}= \frac{N\sigma }{r^2}
 \Bigg [
 \left (1-\frac{2\tau V'^2}{\sigma^2} \right)
  \left (3w^2-1+\frac{8ww'}{r}(w^2-1)-\frac{4w'^2(1-w^2)^2}{r^2} \right)
 \\
 \nonumber
 &&
 +4w'^2\left(\frac{2(1-w^2)^2}{r^2}+\frac{r^2V'^2}{\sigma^2}-1\right)
 +\frac{32\tau^2w^2(1-w^2)^2V'^2}{\sigma^2 r^4(1+\frac{2\tau}{r^4}(1-w^2)^2)}
\Bigg ],
\end{eqnarray}
which is a regular function in the entire range $-\infty<\rho<\infty$.
The corresponding analysis for the odd sector is much more evolved. After much algebra,
the perturbation equations for
$\delta \tilde w(r,t)=\tilde w_1(r) e^{-i\Omega t}$ and $\delta \nu(r,t)=  \nu_1(r) e^{-i\Omega t}$
can be cast in the form
\begin{eqnarray}
\label{schr-odd}
-\frac{d^2 \Psi}{d \rho^{*2}}+U_{odd}(\rho^*)\Psi= \Omega^2 \Psi,
\end{eqnarray} 
where $\Psi=[(1+6\tau (1-w^2)^2/r^4)\nu_1-12 \tau(w^2-1)w'/r^4 \tilde w_1] F_1$,
 $dr/d \rho^*=N \sigma/(1+12 \tau Nw'^2/(r^2(1+6\tau (1-w^2)^2/r^4)))^{1/2} $ 
and $\nu_1(r)=F_2 \Psi+F_3 \Psi'$.
The potential $U_{odd}$ and the functions $F_i$ have complicated expressions 
depending on the equilibrium functions $m,\sigma,w$ and $V$, with  $F_1>0$.
For nodeless solutions, $U_{odd}$ and  $F_i$ are regular in the entire range
$-\infty<\rho^*<\infty$.

For both equations (\ref{schr-even}) and (\ref{schr-odd}),
the potential vanishes near the horizon and at infinity.
Then standard results \cite{Messiah} imply that 
there are no negative eigenvalues for $\Omega^2$ (and hence no unstable modes) 
if the potentials $U_{even}$ and $U_{odd}$  are everywhere positive.

Our results indicate that this is indeed the case for some of the solutions with $\kappa >1/2$,
see $e.g.$  the inlet in Fig. 1. Interestingly, approaching the extremality
appears to imply generically positive values for the potentials in (\ref{schr-even}), (\ref{schr-odd}). 

Therefore we conclude that, in contrast to all other known $d=4$
asymptotically flat hairy black holes with non-Abelian gauge fields only
\cite{Volkov:1998cc}, at least
some of our solutions here are linearly stable.
  
\noindent{\textbf{~~~Further remarks.--~}}
We close with some remarks on the generality of the results in this work.
First, we remark that the $F^2$ and $F^4$ terms in (\ref{action})
are the pieces which  also enter the
Lagrangian of the non-Abelian Born-Infeld theory \cite{Tseytlin:1997csa}
describing the low energy dynamics of $D-$branes.
Although the equations of motion of that model coupled with gravity
are more complicated and do not admit the RN black hole as a solution, 
it is natural to expect that the picture we have found here will share some 
similarities with the results in that case.
 
Non-Abelian fields featuring both the $F^2$ and $F^4$ terms  
appear also in the higher loop corrections to the action of the $d=10$ 
heterotic string \cite{Polchinski:1998rr}.
However, a generalization in that framework of the solutions
considered  here  is not an easy task, due to the occurrence   there  of 
a variety of other fields. 
Previous work in this direction \cite{Donets:1995ya}
indicates that the solutions with a standard $F^2$ term 
only, share the basic features of the EYM black holes in \cite{89}, in particular being unstable.
It
  appears however, that inclusion of higher order gauge field curvature
terms could lead to a very different situation (for example, we have found rather 
similar results to those discussed above, for a  generalization of (\ref{action})
with an extra dilaton field, a Gauss-Bonnet term and gauge group $SO(5)$).
Thus we expect the following picture to be generic:
the $F^4$ term introduces a supplementary interaction between
electric and magnetic fields, which, for some range of the parameters, 
implies a tachyonic mass for the vacuum perturbations of the non-Abelian magnetic fields
around the Abelian solutions. Then the Abelian gauge symmetry is spontaneously 
broken near a black hole horizon for some critical value of the charge to mass ratio,
with the appearance of a condensate of magnetic non-Abelian gauge fields there.
The possible relevance of these aspects in providing analogies to phenomena observed in
condensed matter physics
is yet to be explored.
\\
\\
\noindent{\it{~~Acknowledgements.}}
 This work is carried out in the framework of Science Foundation Ireland (SFI) project
RFP07-330PHY.  
E.R. gratefully acknowledges support by the DFG.


\end{document}